# GnRH induced Phase Synchrony of Coupled Neurons

**Md. Jahoor Alam**

*College of Applied Medical Sciences, University of Ha'il, Kingdom of Saudi Arabia*

**Abstract**

Gonadotropin-releasing hormone (GnRH) is reported to control mammalian reproductive processes. GnRH a neurohormone which is pulsatile released into the pituitary portal blood by hypothalamic GnRH neurons. In the present study, the phase synchronization among a population of identical neurons subjected to a pool of coupling molecules GnRH in extracellular medium via mean-field coupling mechanism is investigated. In the model of populated neurons, GnRH is considered to be autocrine signaling molecule and is taken to be common to all neurons to act as synchronizing agent. The rate of synchrony is estimated qualitatively and quantitatively by measuring phase locking values, time evolution of the phase differences and recurrence plots. Our numerical results show a phase transition like behavior separating the synchronized and desynchronized regimes. We also investigated long range communication or relay information transfer for one dimensional array of such neurons.

## I. INTRODUCTION

The oscillating sharp pulses of Gonadotropin releasing hormone (GnRH) generated by GnRH secreting neurons in the hypothalamus is considered to be an intrinsic property of GnRH secreting neurons and to control important biological functions, for example mammalian reproduction[1,2].The secreted hormone is believed to be responsible for onset puberty and the regulation of hormone release from pituitary[2]. This pulsating behavior of GnRH is done via autofeedback effect of GnRH on its own release via three G-protein types[7] and this GnRH signal is essential for mammalian reproduction because the absence of this signal leads to various reproductive diseases such as sterility etc[2,4]. The sparsely distributed GnRH secreting neurons in the pulse generators are small in number, so the GnRH molecules secreted into the extracellular medium may act as diffusing molecules among the neurons to maintain synchronization of neurons[4-6]. Therefore GnRH has both the roles of feedback regulator as well as synchronizing agent. However, still the way how the cells coupled via GnRH molecules is not fully investigated and understood. We organized our work as follows, in section II, model of the neuron with autocrine signaling mechanism, and method to analyse the coupling of the cells via GnRH signaling molecules. Our numerical results and discussions are explained in section III. Some conclusions based on our numerical results are presented in section IV.

## II. NEURON MODEL AND METHODS

The single cell model developed by Khadra and Li[5] based on experimental data of Krsmanovic[7] was optimized and simplified by Li and Khadra[1] which is schematized in Fig. 20 (A). In this simple model S, Q, I and G denote dimensionless concentration of $\alpha_s$, $\alpha_q$, $\alpha_i$ which are subunits of α and GnRH neurons. $\alpha_s$, $\alpha_q$ and $\alpha_i$ promote secretion of intracellular messengers cAMP and $Ca^{2+}$ ans $IP_3$. The complete model of six variables was simplified based on the idea that two of the six variables are much faster than the other variables and using quasi-steady state approximations. The time evolution of the four variables in the simplified model is given by,

$$\frac{dG}{dt} = \lambda[v + \eta F(S,Q,I) - G] \quad (1)$$

$$\frac{dS}{dt} = \phi[\frac{G^4}{\sigma^4 + G^4} - S] \quad (2)$$

$$\frac{dQ}{dt} = \Psi[\frac{G^4}{\rho^4 + G^2} - Q] \quad (3)$$

$$\frac{dI}{dt} = \varepsilon[\frac{G^2}{\sigma^2 + G^2} - I] \quad (4)$$

where, the parameters are defined as, $F(S,Q,I) = [C_\infty(Q)a_\infty(S,I)]^3$, $C_\infty(Q) = \frac{j_{in}+(\mu+q\delta)c_0}{\mu+1+q\delta}$ and $a_\infty(S,I) = l + \theta S \frac{\omega}{\omega+1}$. $\lambda$, $\phi$, $\Psi$ and $\varepsilon$ are dimensionless parameters characterizing time scale of G, S, Q and I[1]. σ, ρ and k are constants related to inhibition of S, Q and I. $j_{in}$ is the dimensionless rate of $Ca^{2+}$ influx from extracellular medium through voltage gated $Ca^{2+}$ channel. Single cell model is equivalent to a population model of identical neurons[5]. One can consider a population of neurons (heterogeneous) in a common pool of extracellular GnRH as shown in Fig. 2 (B). GnRH in the extracellular medium plays the role of feedback regulator and synchronizing agent[5].

The differential equations in ith neuron is given by, $\frac{dG}{dt} = \frac{1}{N}\sum_{j=1}^{N} \lambda_j [v_j + \eta_j F_j(S_j, Q_j, F_j) - G$ (5)

$$\frac{dS_i}{dt} = \phi_i [\frac{G^4}{\sigma_i^4 + G^4} - S_i] \quad (6)$$

$$\frac{dQ_i}{dt} = \Psi_i [\frac{G^4}{\rho_i^4 + G^2} - Q_i] \quad (7)$$

$$\frac{dI_i}{dt} = \varepsilon_i [\frac{G^2}{k_i^2 + G^2} - I_i] \quad (8)$$

where, i = 1, 2, . . . ,N. The equation (5) indicates the averaging of all contributions from N cells which is similar to mean field coupling. G is taken to be common to all cells. The neurons interact each other by secreting G in the pool and responding to the G in the pool.

The extracellular signal G is shared by the neurons to communicate each other and processing information. When a given variable, say $G_i$ in our case is common to N identical oscillators then this species can be an effective means of mean field coupling, coupling each oscillator to every other via $\frac{1}{N}\sum_{i=1}^{N} G_i$. The common species provides a "mean–field" whereby the oscillators communicate, and the remaining variables synchronize[8,9]. The measure of synchronization of the time evolution of two independent and identical systems can be possible[10] by defining an instantaneous phase for an arbitrary signal $\eta(t)$ via the Hilbert transform[11]

$$\tilde{\eta}(t) = \frac{1}{\pi} P.V \int_{-\infty}^{+\infty} \frac{\eta(t)}{t-\tau} d\tau \quad (9)$$

where P.V. denotes the Cauchy principal value. The instantaneous phase ϕ(t) and amplitude A(t) of a given arbitrary signal can be obtained through the relation, $A(t)e^{i\phi(t)} = \eta(t) + i\tilde{\eta}(t)$. For any given pair of signals [(i, j); i, j = 1, 2, ...,N, I ≠ j], one can therefore obtain the instantaneous phases $\phi_i$ and $\phi_j$ ; phase synchronization is then the condition that $\Delta\phi_{ij} = m\phi_i - n\phi_j$ is constant with m and n being integers. Starting with different initial configurations, the temporal dynamics of the uncoupled oscillators will be uncorrelated; however upon coupling, the dynamics can show phase synchrony[11–13]. Another way to measure the rate of synchrony of two coupled oscillators i$_s$ to plot the two corresponding variables x, x′ from the two oscillators along the two axes of the two dimensional cartesian plane (Pecora-caroll type)[14]. If the oscillators are uncoupled then the points in the plane scattered away from the diagonal. However, if the oscillators are coupled then the points concentrated towards the diagonal. The rate of concentration of the points towards the diagonal measures the rate of synchrony.

### III. RESULTS AND DISCUSSIONS

We present the results of 10 cells out of 50 cells which we have simulated by solving the set of differential equations [5-8] with the values of the parameters used in the simulation in Fig. 2. The upper panel shows the dynamics of I as a function of time in minutes; (i) uncoupled dynamics at k = 454 in time interval (0-2000) minutes (no mean field coupling is swich on), (ii) Synchronized behavior at k = 464 and in time interval (2000-5000) minutes (when mean field coupling is switched on) and (iii) mean field coupling is still switch on but simulated for different values of k = 1 − 1000 after time 5000 minutes. The dynamics in time 〈 5000 minutes, I$_s$ of different cells different peaks which are smaller as the value of k increases but synchronous at foot. To see the synchronized and desynchronized behavior in another way, we show phase plot of the pairs of cells for variable I for all three regions in the lower panel. For uncoupled cells, Δϕ fluctuates randomly in time 〈 2000 minutes, whereas, it becomes constant for synchronized regime i.e. in (2000 〈 time 〈 5000). Furthermore, since Δϕ fluctuates but not randomly in 5000 〈 time 〈 7000, the dynamics mare weakly synchronized.

To see the synchronization of a group of neurons coupled via GnRH, we take N=100 such neurons and coupling is switch on at t=4000 minutes as shown in Fig. 3. The phase plots of all the concentration variables, s, q and i as a function time indicate clearly the desynchronized ($\Delta\phi_j^{mn}$, m, n = 1, 2, 3, ...,N, j = s, q, i fluctuates randomly with time) and synchronized regimes ($\Delta\phi_j^{mn}$ fluctuates about a constant value). Our claim is supported by the pecora-carol type recurrence plots of the respective variables.

## IV. CONCLUSION

In autocrine signaling mechanism which we discussed in GnRH neuron model, not only the cells could able to send signals to other cells of the same type, but also they can send signals to themselves. This way of communication was shown by our simulation results which we have done among a group of cells subjected in a pool of signaling molecule, GnRH and coupling the cells via mean field coupling mechanism.

Since the GnRH molecules can diffuse not only to the cell itself but also can diffuse to other cells. So synchronization of the cells is achieved via this signaling molecule.

# References

1. Y.X. Li and A. Khadra, *Bull. Math. Biol.* 70, 2103-2125 (**2008**).

2. A. Khadra, *Physica D: Nonlinear Pheomena* 238,771 (**2009**).

3. L.Z. Krsmanovic, N. Mores, C.E. Navarro, K.K. Arora and K.J. Catt, *Proc. Natl. Acad. Sci.* 100, 2969-2974 (**2003**).

4. T.A. Richter and E. Terasawa, *Trends Endocrinol. Metab*. 12, 353-359 (**2001**).

5.  A. Khadra and Y.X. Li, *Biophys. J* 91, 74-83 (**2006**).

6. Liang Cheng, Mingwang Shao, Mingliang Zhang, and Dorothy Duo Duo Ma, *Sci. Adv. Mater.* 2, 386-389 (**2010**) .

7. L.Z. Krsmanovic, N. Mores, C.E. Navarro, K.K. Arora and K.J. Catt, *Proc. Natl. Acad. Sci*. 100, 2969-2974 (**2003**).

8. D. Gonze, J. Halloy, JC. Leloup, A. Goldbeter, *C. R. Biol.*, 326, 189 (**2003**).

9. M. G. Rosenblum and A. S. Pikovsky, *Phys. Rev. Lett.*, 92, 114102 (**2004**).

10. H. Sakaguchi and Y. Kuramoto, *Prog. Theor. Phys.*, 76, 576 (**1986**).

11. M. G. Rosenblum, A. S. Pikovsky and J. Kurths, *Phys. Rev. Lett*., 76, 1804 (**1996**).

12. A. Pikovsky, M. Rosenblum and J. Kurths, *Synchronization: A Universal Concept in Nonlinear Science (Cambridge University Press, Cambridge,* (**2001**).

13. A. Nandi, Santhosh G., R. K. B. Singh and R. Ramaswamy, *Phys. Rev. E,* 76, 041136 (**2007**).

14.  L. M. Pecora and T. L. Caroll, *Phys. Rev. Lett.,* 64, 821 (**1990**).

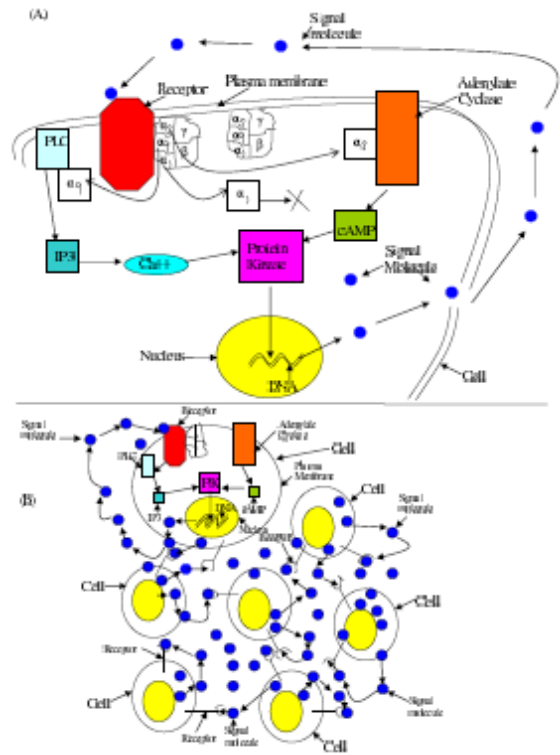

**FIG. 1**: (A) The schematic diagram of molecular mechanism of the autocrine signaling in the model and (B) Coupling of a group of cells in a pool of GnRH.

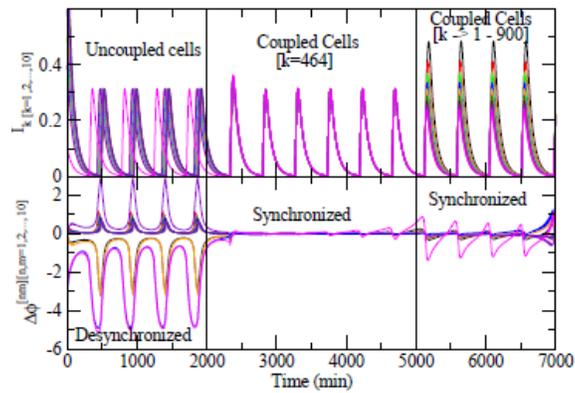

**FIG. 2**: Plots of $I_j$, (j=1,2,...,10) of 10 cells with common GnRH (G) at k = 464 showing desynchronized, synchronized and sparsely synchronized regimes. The values of the parameters we have taken in the simulation are: l = 0.067, $j_{in}$ = 9.227 × $10^{-6}$, m = 0.012, $\tilde{c}$ = 0.588, $c_0$ = 0.588, $a_l$ = 1.0, θ = 216.67, $\omega$ = 0.01125, n = 0.706, η = 3.292, $\phi$ = 1.0, σ = 1.0, ψ = 1.0, ρ = 61.765, $\varepsilon$ = 0.0125 and k = 1.0.

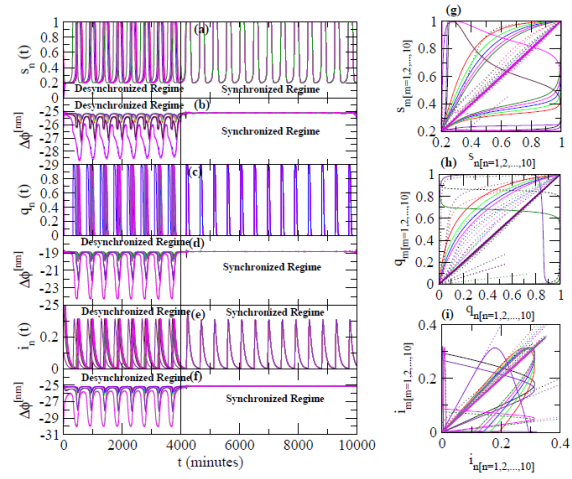

**FIG. 3:** Plots of $s_j$, $q_j$ and $i_j$ concentration variables, (j=1,2,...,10) of 10 cells with common GnRH (G) at k = 500 showing desynchronized and synchronized regimes in panels (a), (c) and (e) respectively. The corresponding phase plots of the pairs of the cells of the variables are shown in panels (b), (d) and (f). Similarly, the pecora-carol type recurrence plots of all variables are in the panels (g), (h) and (i) respectively. The values of the parameters we have taken in the simulation are the same as in Fig. 2.